# *Ab initio* simulation of Ta$_2$O$_5$: A high symmetry ground state phase with application to interface calculation

Jun-Hui Yuan,[1] Kan-Hao Xue,[1,*] Qi Chen,[1] Leonardo R. C. Fonseca,[2] and Xiang-Shui Miao[1, #]

[1] Wuhan National Research Center for Optoelectronics, School of Optical and Electronic Information, Huazhong University of Science and Technology, Wuhan 430074, China

[2] Departamento de Física, ICEx, Universidade Federal de Minas Gerais, 30123-970 Belo Horizonte, MG, Brazil

## Corresponding Authors

*E-mail: xkh@hust.edu.cn (K.-H. Xue)     #E-mail: miaoxs@hust.edu.cn (X.-S. Miao)

## ABSTRACT

We suggest a tetragonal *I*4$_1$/*amd* phase (η-phase) as the ground state of Ta$_2$O$_5$ at zero temperature, which is a high symmetry version of the triclinic γ-phase Ta$_2$O$_5$ predicted by Yang and Kawazoe. Our calculation shows that γ-phase Ta$_2$O$_5$ will automatically be transformed into the η-phase during structural relaxation. Phonon dispersion confirms that the η-phase is dynamically stable, while the high temperature α-phase Ta$_2$O$_5$, which also has the *I*4$_1$/*amd* symmetry, is unstable at zero temperature. A thorough energy comparison of the β$_{AL}$, δ, λ, B, L$_{SR}$, β$_R$, *Pm*, *Cmmm*, γ, η and α phases of Ta$_2$O$_5$ is carried out. The GGA-1/2 method is applied in calculating the electronic structure of various phases, where the η-phase demonstrates a 4.24 eV indirect band gap, close to experimental value. The high symmetry tetragonal phase together with computationally efficient GGA-1/2 method greatly facilitate the *ab initio* simulation of Ta$_2$O$_5$-based devices. As an example, we have explicitly shown the Ohmic contact nature between metal Ta and Ta$_2$O$_5$ by calculating an interface model of *b.c.c.* Ta and η-Ta$_2$O$_5$, using GGA-1/2.

# I. INTRODUCTION

Tantalum penta-oxide ($Ta_2O_5$) is a versatile functional dielectric that can be used as electrolytic capacitor,[1] optical modulator,[2] optical coating,[3] high-K dielectric material in microelectronics,[4] and the switching layer in memristors.[5]–[10] Despite its wide range of applications, the knowledge of its exact structure is nevertheless still limited. At zero temperature, the structure of ground state $Ta_2O_5$ crystal has been under debate for decades.[11]–[13] On the other hand, the applied computational researches on $Ta_2O_5/TaO_x$ have been much more rare[14],[15] compared with $HfO_2/HfO_x$,[16]–[27] though $TaO_x$ and $HfO_x$ share many properties in common and both are regarded as the most promising candidates for next-generation memristors. This state of affairs is strongly related to the uncertainty regarding which model to use for the ground state of $Ta_2O_5$. Indeed, most computational works have been still focusing on revealing the ground state atomic structures of $Ta_2O_5$.[13] Furthermore, to correctly describe the electronic structure of various $Ta_2O_5$ phases remains a challenging task, where one either accepts the band gap underestimation stemming from the local density approximation[28] (LDA) and generalized gradient approximation[29],[30] (GGA) of density functional theory[31] (DFT), or has to resort to hybrid functionals[32] and quasi-particle approaches such as the $GW$ approximation.[33] Although Cui and Jiang obtained excellent $GW$ electronic structures for $Ta_2O_5$,[34] the computational load is formidable when large supercells involving interfaces and point defects are considered. In order to investigate complex structures at the device level, it is necessary to develop a more efficient methodology for the theoretical simulation of large $Ta_2O_5$-based supercells involving hundreds of atoms.

A comparison with $HfO_2$ further clarifies the difficulty regarding the electronic structure calculation for $Ta_2O_5$. While DFT-LDA and DFT-GGA also severely underestimate the band gap of $HfO_2$, the more accurate $G_0W_0$ method reveals relatively uniform band gap values among its various phases, such as monoclinic ($P2_1/c$, 5.45 eV), tetragonal ($P4_2/nmc$, 5.78 eV), and cubic ($Fm\overline{3}m$, 4.91 eV),[35] thus one may calculate any hafnia system using conventional GGA while the true band gap can readily be estimated with good reliability. On the contrary, the calculated band gap values for various $Ta_2O_5$ phases differ a lot. For instance, using $GW$ approximation Lee $et\ al.$ reported 1.03 eV, 2.22 eV, 2.96 eV and 4.26 eV band gaps for β-$Ta_2O_5$ ($Pccm$), δ-$Ta_2O_5$, orthorhombic $Ta_{22}O_{55}$ and amorphous $Ta_2O_5$, respectively.[36] Moreover, while the primitive cells of various $HfO_2$ phases contain few atoms, typically not more than 12, many models for $Ta_2O_5$ require large primitive cells, even containing 77 atoms in the famous model of Stephenson and Roth.[11] Very recently, Yang and Kawazoe proposed a new triclinic phase for $Ta_2O_5$,[13] whose energy is lower than any other proposed $Ta_2O_5$ models. However, due to the lack of symmetry, its adoption in interfacial models is quite challenging.

In this work, we suggest a tetragonal $I4_1/amd$ phase (referred to as η phase) as a prototype ground state of $Ta_2O_5$, which can be regarded as a high symmetry counterpart of the triclinic structure proposed by Yang and Kawazoe. The self-energy corrected LDA-1/2 and GGA-1/2 methods (referred to as DFT-1/2 in general) are utilized to calculate the electronic structures of $\eta$-$Ta_2O_5$, recovering band gaps close to experimental value. The new high symmetry phase and the efficient DFT-1/2 method can be combined to investigate several practical issues relevant to $Ta_2O_5$-based devices.

## II.    COMPUTATIONAL

### Existing model structures for $Ta_2O_5$

As our aim is to identify the most suitable $Ta_2O_5$ phase for ground state density functional calculations, we shall focus here on the low temperature phases of $Ta_2O_5$ (L-$Ta_2O_5$). Yet, knowledge of the high temperature phases is still valuable since they are usually more symmetric. Above $1360^oC$,[37] $Ta_2O_5$ is in the high temperature phase named H-$Ta_2O_5$ or $\alpha$-$Ta_2O_5$, for which various model structures have been proposed including triclinic, monoclinic, orthorhombic and tetragonal structures.[38],[39] Fortunately, the most symmetric tetragonal phase has been fully refined, in an $I4_1/amd$ symmetry.[39]

For L-$Ta_2O_5$, early in the 1960s, hexagonal (named $\delta$-$Ta_2O_5$ or TT-$Ta_2O_5$), monoclinic and orthorhombic (usually referred to as $\beta$-$Ta_2O_5$) phases of L-$Ta_2O_5$ were already reported experimentally.[40]–[42] Based on X-ray diffraction (XRD) patterns, Stephenson and Roth initially proposed an orthorhombic model for L-$Ta_2O_5$ in 1971, with a unit cell of $Ta_{22}O_{55}$.[11] This model, usually named as the $L_{SR}$ model in the literature, contains a certain amount of oxygen vacancies to reach the $Ta_2O_5$ stoichiometry. Several years later, a high pressure phase of L-$Ta_2O_5$ was discovered by Izumi and Kodama through hydrothermal synthesis,[43] referred to as B-$Ta_2O_5$. It is isomorphic with the B-$Nb_2O_5$, both possessing a monoclinic $C2/c$ structure. However, B-$Ta_2O_5$ was found to be stable even at atmospheric pressure and low temperature.[43],[44] In 1992, Hummel *et al.* identified the space group of δ-$Ta_2O_5$ as $P6/mmm$, but the exact atomic structure was still unknown.[45] Meanwhile, starting from δ-$Ta_2O_5$ they also synthesized a new single crystal phase of $Ta_2O_5$, the so-called T-$Ta_2O_5$ with an orthorhombic $Pmm2$ symmetry. The exact structure was identified, which actually has the non-stoichiometric $Ta_{24}O_{62}$ composition.[45] Later in 1997, Fukumoto and Miwa identified the exact atomic structure of δ-$Ta_2O_5$ using first-principles calculations,[46] which has a $Ta_4O_{10}$ unit cell with the $P6/mmm$ symmetry.

Research into the atomic structure of L-$Ta_2O_5$ has been very intensive in recent years, due to the importance of $Ta_2O_5$ in microelectronics. Sawada and Kawakami proposed a simplified $Ta_6O_{15}$ model supercell instead of the $L_{RS}$ model in 1999,[47] so as to facilitate the simulation. This model relaxes to a monoclinic structure with $Pm$ symmetry in our calculation. Through Rietveld full-profile analysis on the XRD patterns, Aleshina and Loginova obtained a $Pccm$ model structure for orthorhombic L-$Ta_2O_5$ in 2002,[48] denoted as the $β_{AL}$ phase. Shortly after that, Ramprasad proposed another orthorhombic model with a $Ta_4O_{10}$ unit cell (denoted as $β_R$ model in the literature),[49] which was an simplified version of the $L_{SR}$ model for the sake of simulation efficiency. The original $\beta_R$ model has a *Pm* symmetry, but after relaxation it becomes a more symmetric *Pmma* phase in our calculation. In 2005, Grey *et al.* synthesized a new phase of single crystal L-$Ta_2O_5$ in the monoclinic *C2/m* symmetry, whose unit cell composition is $Ta_{38}O_{95}$. In 2013, Lee *et al.* predicted a high symmetry *Pbam* phase for orthorhombic $Ta_2O_5$ with a simple $Ta_4O_{10}$ unit cell, which is called $\lambda$-$Ta_2O_5$.[50] The energy of $\lambda$-$Ta_2O_5$ is, according to the original calculation, lower than the $L_{SR}$ model, but it remained uncertain whether it is the most suitable ground state for L-$Ta_2O_5$. Later, an orthorhombic *Cmmm* structure was proposed by Kim *et al.* in 2014,[51] which is energetically more favorable than $\lambda$-$Ta_2O_5$ according to their calculation.

Most recently, Yang and Kawazoe identified a triclinic phase of $Ta_2O_5$ using *ab initio* evolutionary algorithm, as a new ground state for L-$Ta_2O_5$.[13] Named as $\gamma$-$Ta_2O_5$, this phase was reported to be considerably more stable than any other known phase of $Ta_2O_5$ at zero temperature and atmospheric pressure, though the *Cmmm*-phase $Ta_2O_5$ was not explicitly listed in their comparison. The discovery of the $\gamma$-phase can be regarded as a great success of modern structural prediction methodology by combining density functional theory and the evolutionary algorithm. Yet, the triclinic nature of this phase may impose certain difficulties in model construction for $Ta_2O_5$-related interfaces and capacitors. Therefore, it is still worthwhile to develop a simple orthorhombic (or even more symmetric) phase for L-$Ta_2O_5$, whose energy is lower than $\lambda$-$Ta_2O_5$ and close to $\gamma$-$Ta_2O_5$.

## *Fundamental computational settings*

DFT calculations were carried out using the plane-wave-based Vienna *Ab initio* Simulation Package,[52],[53] with a fixed 500 eV plane-wave kinetic energy cutoff. For the exchange-correlation (XC) energy, three distinct functionals were considered: the LDA functional of Ceperley-Alder[54] parameterized by Perdew and Zunger[55] (CA), the GGA functional of Perdew-Burke-Ernzerhof (PBE),[30] and the PBEsol functional that was specially designed for solids.[56] The electrons considered as valence were: $5p$, $5d$ and $6s$ for Ta; $2s$ and $2p$ for O. Core electrons were approximated by projector augmented-wave pseudopotentials.[57],[58] In all self-consistent runs, the convergence criterion for total energy was $10^{-6}$ eV, while structural optimization was reached for Hellmann-Feynman forces on any atom less than 0.01 eV/Å in each direction.

## *DFT-1/2 for improved band structures*

The band gap underestimation by DFT-LDA and DFT-GGA has long been a well-known problem,[59] which is attributed to the lacking of derivative discontinuity in the XC functionals.[60],[61] The DFT-1/2 method, proposed by Ferreira *et al.* in 2008,[62] attributed the band gap underestimation to a self-energy term of the valence band hole in the intrinsic excitation of a semiconductor. The subtraction of this self-energy rectifies the band gap, but DFT-1/2 does such correction within the Kohn-Sham framework, unlike the *GW* approximation that goes beyond.

In practice, one does not calculate the self-energy of the hole using perturbative methods. Rather, since the hole is localized near the anions, one simply attaches the corresponding "self-energy potential" $V_S$ to these anions (usually summed to the pseudopotential) in real space. The self-energy correction is achieved after a second self-consistent calculation using the self-energy corrected pseudopotential. To keep the self-energy potential local to the corresponding anion, it must be trimmed before summing to anion pseudopotential. The optimal cutoff radius is obtained variationally by maximizing the band gap. Because the self-energy potentials are introduced in the solid state calculations as external atomic potentials, the ground state energies given by DFT-1/2 self-consistent calculations are not physically meaningful. Nevertheless, electronic band diagrams and density of states are properly corrected by DFT-1/2.[62] Hence, in our work all total energies were taken from DFT calculations, while the band diagrams were obtained from either DFT or DFT-1/2 calculations.

Since $Ta_2O_5$ is a binary metal oxide, the self-energy correction ought to be carried out only for the oxygen anions. The optimal cutoff radius to trim the corresponding $V_S$ (with half electron subtracted from the O 2p orbital[62]) was obtained variationally upon maximizing the band gap,[62] and no empirical parameters were involved.

The oxygen self-energy potentials specific to all these XC functionals (LDA-CA, GGA-PBE, GGA-PBEsol) were derived from atomic calculations using a modified ATOM code (supplied with the Siesta simulation package[63]). We have also created a web-based self-energy correction program, where pseudopotentials modified by the inclusion of DFT-1/2 self-energy potentials can be generated online for all these XC functionals.[64]

## III.   RESULTS AND DISCUSSIONS

### *Optimized crystal structures of previously identified L-Ta$_2$O$_5$ phases*

We studied in detail nine representative L-Ta$_2$O$_5$ phases, as shown in **Figure 1**. The *Pmm*2 phase was not considered because it is non-stoichiometric, while the *C*2/*m* phase was omitted because of its rather complex unit cell (Ta$_{38}$O$_{95}$). In addition, the high temperature $\alpha$-phase was also included for comparison. The optimized lattice parameters for these ten structures are summarized in **Table 1**, together with the corresponding theoretical and experimental values in the literature.

**Table 1** Calculated structural parameters for L-Ta$_2$O$_5$ in $\beta_{AL}$, $\delta$, $\lambda$, B, L$_{SR}$, $\beta_R$, *Pm*, *Cmmm*, $\gamma$ and $\alpha$ phases using various XC functionals. Previous reported theoretical and experimental values are listed for comparison. In parenthesis we indicate the relative error with respect to the experimental value.

| | LDA | PBE | PBEsol | Literature | |
| --- | --- | --- | --- | --- | --- |
| | | | | Theory | Expt. |
| | | | $\beta_{AL}$-phase | | |
| a (Å) | 6.371(+2.45%) | 6.511(+4.71%) | 6.424(+3.33%) | 6.52[13] | 6.217[48] |
| b (Å) | 3.658(-0.52%) | 3.698(+0.57%) | 3.676(-0.03%) | 3.69 | 3.677 |
| c (Å) | 7.691(-1.32%) | 7.778(-0.21%) | 7.728(-0.85%) | 7.78 | 7.794 |
| | | | $\delta$-phase | | |
| a (Å) | 7.215(-0.34%) | 7.338(+1.35%) | 7.267(+0.37%) | 7.33[13] | 7.24[41] |
| c (Å) | 3.845(-0.90%) | 3.888(-0.21%) | 3.863(-0.44%) | 3.89 | 3.88 |
| $\gamma$ angle (°) | 120 | 120 | 120 | 120 | |

| | | | | λ-phase | |
|---|---|---|---|---|---|
| a (Å) | 6.160 | 6.258 | 6.199 | 6.25[13] | |
| b (Å) | 7.270 | 7.412 | 7.332 | 7.40 | |
| c (Å) | 3.770 | 3.826 | 3.795 | 3.82 | |
| | | | | B-phase | |
| a (Å) | 12.759(-0.20%) | 12.941(+1.22%) | 12.834(+0.38%) | 12.93[13] | 12.785[65] |
| b (Å) | 4.824(-0.62%) | 4.922(+1.40%) | 4.867(+0.27%) | 4.92 | 4.854 |
| c (Å) | 5.484(-0.80%) | 5.593(+1.17%) | 5.528(0.00%) | 5.59 | 5.528 |
| β angle (°) | 104.22(-0.04%) | 103.20(-1.02%) | 103.77(-0.47%) | 103.23 | 104.26 |
| | | | | $L_{SR}$-phase | |
| a (Å) | 6.230 (+0.52%) | 6.336 (+2.23%) | 6.274 (+1.23%) | 6.33[13] | 6.198[11] |
| b (Å) | 40.223 (-0.17%) | 41.058 (+1.91%) | 40.627 (-0.84%) | 40.92 | 40.290 |
| c (Å) | 3.787 (-2.60%) | 3.843 (-1.16%) | 3.812 (-1.95%) | 3.85 | 3.888 |
| γ angle (°) | 90.12 (+0.13%) | 90.13 (+0.14%) | 89.97 (-0.03%) | 89.16 | 90 |
| | | | | $β_R$-phase | |
| a (Å) | 6.141 | 6.221 | 6.173 | 6.03[49] | |
| b (Å) | 7.244 | 7.431 | 7.322 | 7.13 | |
| c (Å) | 3.831 | 3.874 | 3.850 | 3.82 | |
| | | | | $Pm$-$Ta_6O_{15}$ | |
| a (Å) | 6.241 | 6.372 | 6.295 | | |
| b (Å) | 3.759 | 3.822 | 3.789 | | |
| c (Å) | 11.058 | 11.246 | 11.139 | | |
| β angle (°) | 87.95 | 87.32 | 87.61 | | |
| | | | | $Cmmm$-phase | |
| a (Å) | 3.818 | 3.871 | 3.845 | 3.79[51] | |
| b (Å) | 12.937 | 13.146 | 13.018 | 12.79 | |
| c (Å) | 3.845 | 3.891 | 3.865 | 3.81 | |
| | | | | γ-phase | |
| a (Å) | 3.836 | 3.883 | 3.858 | 3.89[13] | |
| b (Å) | 3.835 | 3.884 | 3.858 | 3.89 | |
| c (Å) | 13.209 | 13.421 | 13.294 | 13.38 | |
| α angle (°) | 81.708 | 81.710 | 81.645 | 81.77 | |
| β angle (°) | 98.349 | 98.336 | 98.332 | 98.25 | |
| γ angle (°) | 90.001 | 90.001 | 90.000 | 89.67 | |
| | | | | α-phase | |
| a (Å) | 3.893 (+0.85%) | 3.872 (+0.31%) | 3.834 (-0.67%) | 3.85[66] | 3.86[39] |
| c (Å) | 36.450 (+0.75%) | 36.573 (+1.09%) | 36.031 (-0.41%) | 37.45 | 36.18 |

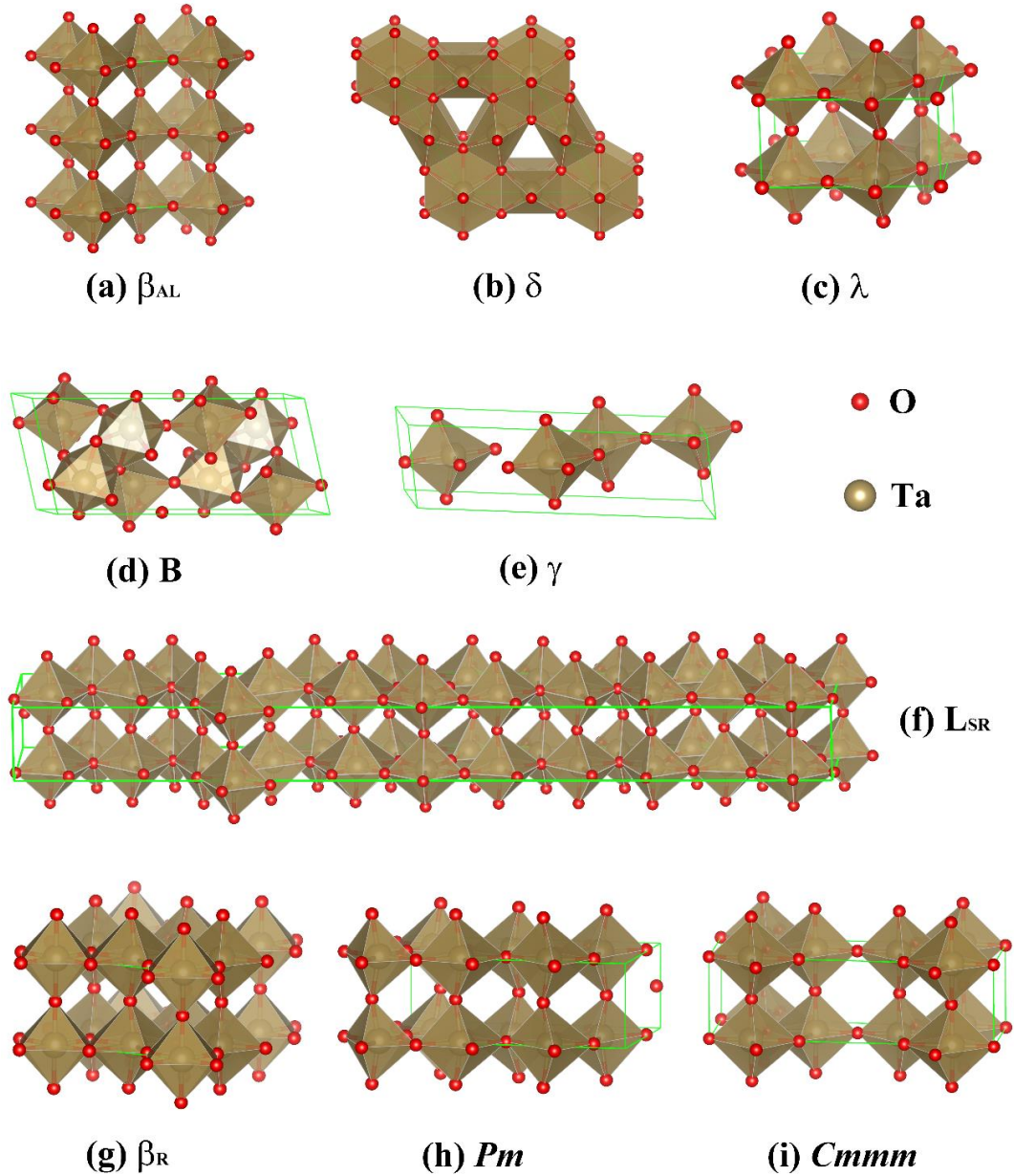

**Figure 1.** Crystal structures of various $Ta_2O_5$ models/phases: (a) $\beta_{AL}$, (b) $\delta$, (c) $\lambda$, (d) B, (e) $\gamma$, (f) $L_{SR}$ , (g) $\beta_R$, (h) *Pm*, and (i) *Cmmm*.

Our results are in general consistent with the data reported in the literature, with the $\gamma$-phase as the only exception. While the exact space group was not given in the original work, the structural data literally point to a $P1$ space group with the lowest symmetry. We thus carried out structural relaxations on the $\gamma$-phase $Ta_2O_5$ in the $P1$ symmetry.

Remarkably, our γ angle is 90º within the computational error range, regardless of the XC functional used. This implies that the γ phase may in fact possess higher symmetry which makes it worthwhile to carry out an in-depth investigation of this issue.

## *A high-symmetry model structure for the ground state of L-Ta$_2$O$_5$*

As discussed above, our optimization of the γ-phase accompanies a change of the γ-angle from 89.67º in the original work[13] to 90.00º in the present work. Since the total energy gets lower during structural relaxation, this implies that the true ground state can be assigned a more symmetric space group. Confined to the *P*1 (No. 1) symmetry, we had missed certain symmetry elements because no calculation could yield absolute accuracy. A careful inspection on our relaxed "γ-phase" led to a much higher symmetry in the space group *I*4$_1$/*amd* (No. 141), whose full structural data are listed in **Table 2**.

**Table 2** Structural parameters of η-Ta$_2$O$_5$ in the *I*4$_1$/*amd* symmetry.

| | Space group | | a (Å) | c (Å) |
|---|---|---|---|---|
| η-Ta$_2$O$_5$ | *I*4$_1$/*amd* | | 3.883 | 26.272 |
| | | | | |
| Sites | x | y | z | Wyckoff position |
| Ta 1 | 0 | 0 | 0.0729 | 8e |
| O 1 | 0 | 0 | 0 | 4a |
| O 2 | 0 | 0 | 0.1570 | 8e |
| O 3 | 0 | 0 | 0.6769 | 8e |

Since the γ-phase was already assigned the triclinic symmetry, we here refer to this high symmetry *I*4$_1$/*amd* phase as η-phase. To demonstrate their relations, in **Figures 2(a)** and **2(b)** we plot 3×3×3 supercells of the γ-phase and the η-phase for comparison. First we notice that the *a* and *b* axes are orthogonal in the η-phase, but not in the γ-phase. Secondly, the marked O-Ta-O bond angle (highlighted in **Figure 2(a)**) along the *b*-axis is 171.5º in γ-Ta$_2$O$_5$, while in η-Ta$_2$O$_5$ it is 180.0º. The high symmetry of the η-phase is not clear when viewed along the *c*-direction (**Figure 2(c)**), because the *c*-axis is not orthogonal to the *a-b* plane. Yet, the tetragonal symmetry is manifested when viewing along the *z*-direction, which is exactly the direction orthogonal to the *a-b* plane (**Figure 2(d)**).

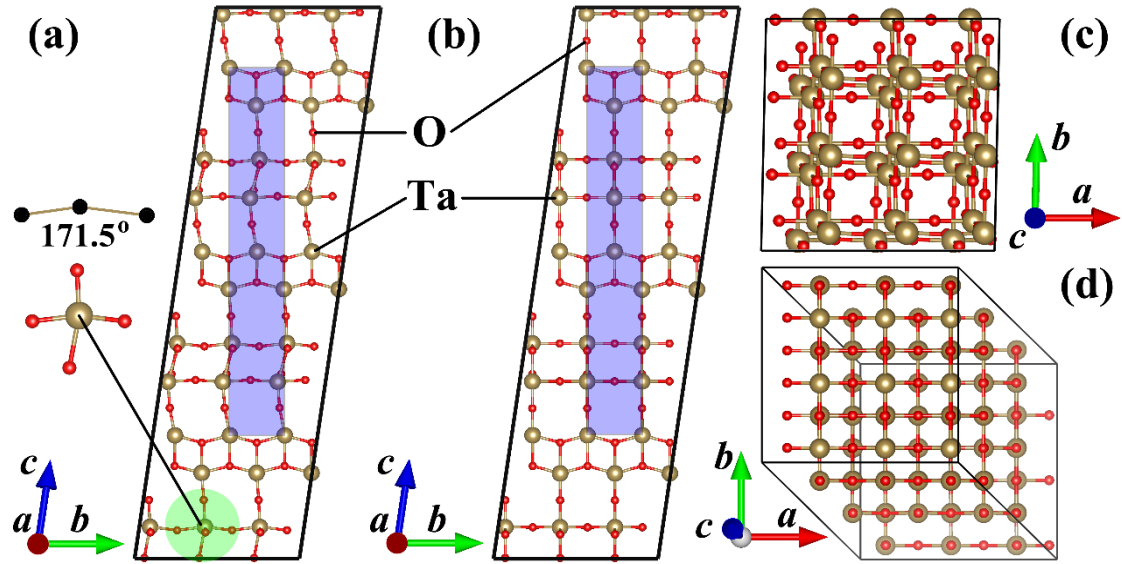

**Figure 2.** (a) Atomic structure of the original γ-phase Ta$_2$O$_5$; (b) fully relaxed "γ-phase" in our calculation, viewed along the *a*-xis; (c) view of the structure in (b) along the *c*-axis; (d) view of the structure in (b) along *z*-axis, which is orthogonal to the *a-b* plane.

The tetragonal unit cell of η-Ta$_2$O$_5$ can be extracted from the rectangular shaded region in **Figure 2(b)**. Interestingly, the most widely accepted phase of H-Ta$_2$O$_5$ (α-Ta$_2$O$_5$) also has the *I*4$_1$/*amd* symmetry. It is a very unusual situation that the zero temperature phase and the high temperature phase share the same space group. Therefore, we make a structural comparison of the $I4_1/amd$ L-$Ta_2O_5$ and $I4_1/amd$ H-$Ta_2O_5$ in **Figure 3**. The α-$Ta_2O_5$ structure is more complicated in that it involves fractional (75%) occupation O sites. Besides, the Ta-O bonds in η-$Ta_2O_5$ mainly extend along *a*-, *b*- and *c*-axes, with only slight distortions in certain regions. Nevertheless, the Ta-O-Ta bonds at the fractional occupation O sites make big angles with the axes, on the *a-b* and *a-c* planes. The existence of oxygen vacancies in α-$Ta_2O_5$ reduces the Gibbs free energy through the entropy-related term at finite temperatures.

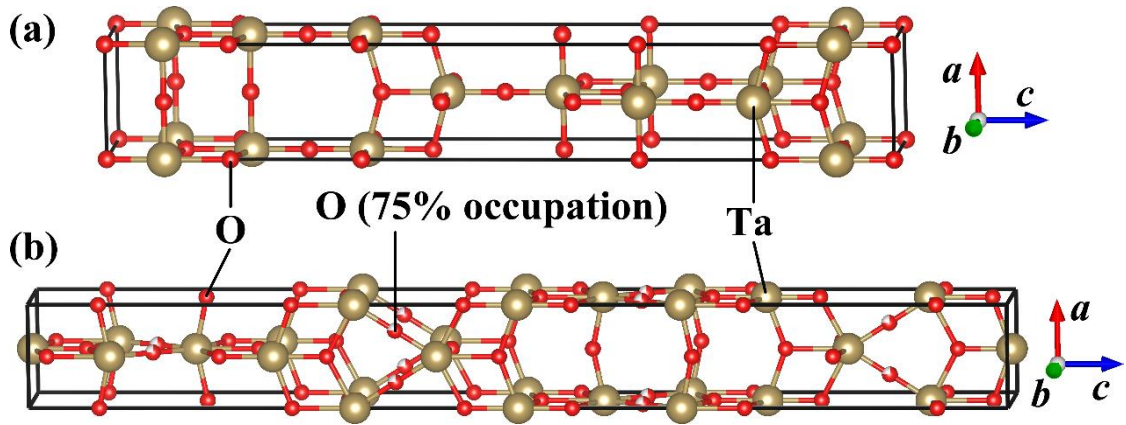

**Figure 3.** (a) Atomic structure of the $I4_1/amd$ η-phase $Ta_2O_5$; (b) atomic structure of the $I4_1/amd$ H-$Ta_2O_5$ (α-$Ta_2O_5$).

Since the γ-phase automatically transforms into the η-phase in our structural optimization, we compared the energies of the two phases in a special way. We fixed the shape of γ-$Ta_2O_5$ to the original settings of Yang and Kawazoe,[13] and optimized the cell volume and atomic coordinates, in order to keep the triclinic symmetry. In this way we find that the η-phase is energetically lower than the γ-phase by 0.0016 eV per $Ta_2O_5$ chemical formula. Compared with other phases, we confirm that the η-phase has the lowest energy among various $Ta_2O_5$ models at zero temperature, where the relative energetics obtained with GGA(PBE) are indicated in **Figure 4**. The reference zero energy was taken as that of the relaxed a-$Ta_2O_5$ model shown in **Figure 5(a)**. The energetics results indicate that most of the recently proposed orthorhombic (or with slight monoclinic distortion in the case of *Pm* phase) models are lower in energy then the amorphous phase. Actually, the arrangement of the Ta-O octehedra is very similar among $L_{SR}$, $\beta_R$, *Pm* and *Cmmm* phases, as seen in **Figure 1**. The two lowest energy phases are obviously the η-phase and the γ-phase. The *Cmmm* phase is the third lowest one in energy, whose energy per $Ta_2O_5$ formula is higher than the η-phase by merely 0.09 eV. Finally, we notice that the high temperature α-$Ta_2O_5$ phase is slightly higher in energy than the amorphous phase, by 0.17 eV per $Ta_2O_5$ formula.

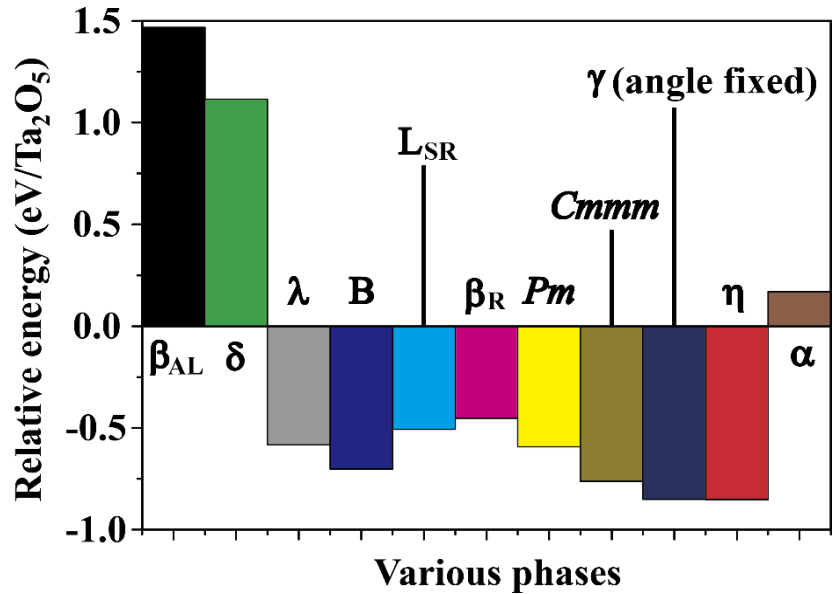

**Figure 4.** Relative energies per unit formula for $Ta_2O_5$ in various phases, calculated using GGA(PBE). The energy of the amorphous phase (a-$Ta_2O_5$) is set to zero.

The dynamic stability of η-$Ta_2O_5$ was examined through its phonon spectra. As shown in **Figure 6**, no imaginary frequency modes are observed, indicating dynamic robustness. Due to its high symmetry, low energy and dynamically stable nature, η-Ta$_2$O$_5$ is highly recommended as the ground state phase to be used in *ab initio* simulation for Ta$_2$O$_5$.

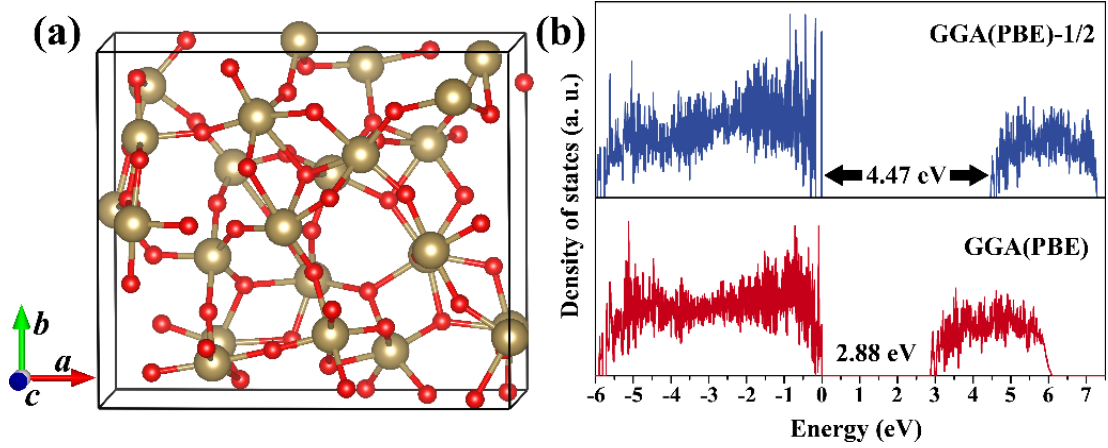

**Figure 5.** (a) Our optimized amorphous Ta$_2$O$_5$ (a-Ta$_2$O$_5$) model; (b) its density of states calculated by GGA(PBE) and GGA(PBE)-1/2.

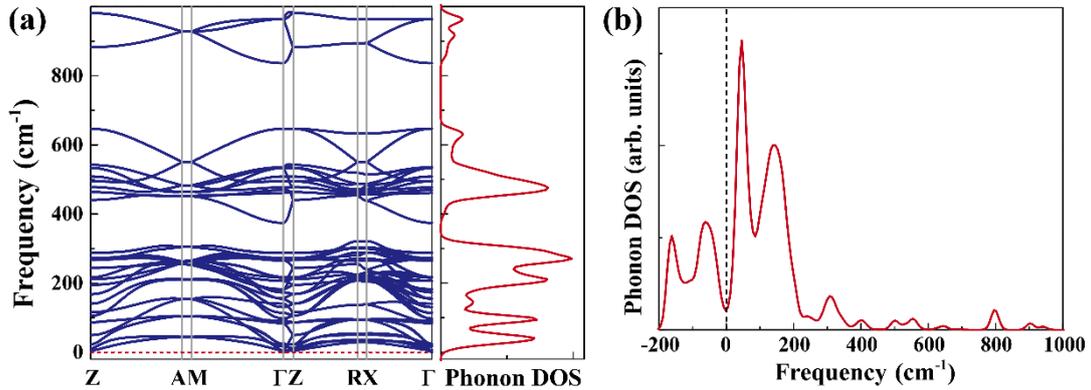

**Figure 6.** (a) Calculated phonon dispersion and phonon density of states of η-Ta$_2$O$_5$ at zero temperature; (b) calculated phonon density of states of α-Ta$_2$O$_5$ (H-Ta$_2$O$_5$) at zero temperature.

We finally note that η-Ta$_2$O$_5$ is actually isomorphic to the tetragonal phase P-Nb$_2$O$_5$, whose structure was first revealed by Petter and Laves in 1965,[67] with a = 3.90 Å and c = 25.43 Å. Initially, Petter and Laves named this phase η-Nb$_2$O$_5$, therefore we refer to the $I4_1/amd$ phase of $Ta_2O_5$ as η-$Ta_2O_5$ following this convention. Yet, researches on η-$Nb_2O_5$ have been rare compared with the more ordinary form of monoclinic H-$Nb_2O_5$. Laves, Moser and Petter observed the similarity of η-$Nb_2O_5$ to α-$Ta_2O_5$, and suggested that the space group of η-$Nb_2O_5$ may be $I4_1$ or $I4_122$.[68] Several later works by Valencia-Balvín *et al.* [69] and Xu et al.[70] assigned the space group $I4_122$ (No. 98, point group $D_4$) to η-$Nb_2O_5$. However, our calculation reveals that the prototype η-$Nb_2O_5$ has the $D_{4h}$ point group and belongs to a higher symmetry space group $I4_1/amd$, identical to that of η-$Ta_2O_5$.

### *Electronic structures of η-$Ta_2O_5$ with comparison to other phases*

The experimental band gap value of L-$Ta_2O_5$ ranges from 3.9 eV to 4.5 eV, [71]–[73] usually measured in thin film format where the band gap can be larger than in the bulk. A reasonable calculated band gap value for bulk L-$Ta_2O_5$ should be slightly greater than 4 eV. Our calculated band gap for a-$Ta_2O_5$, which typically possesses larger band gap than crystalline phases, is 2.88 eV using GGA(PBE) (lower part of **Figure 5(b)**). Such severe underestimation is expected due to the missing of self-energy correction. To carry out self-energy correction using GGA-1/2, we attached the trimmed self-energy potential of oxygen to a-$Ta_2O_5$, using various cutoff radii with a power index of 20 in the cutoff function.[64] The optimum cutoff radius was fixed to 2.3 bohr through a variation process, which eventually maximized the band gap. The corresponding electronic density of states for a-$Ta_2O_5$, as shown in the upper part of **Figure 5(b)**, yields a GGA-1/2 band gap of 4.47 eV, which is consistent with the experimental values for

a-Ta$_2$O$_5$ (4.2 eV to 4.5 eV[73]).

**Table 2** Calculated DFT and DFT-1/2 band gaps (eV) for (from top to bottom) β$_{AL}$, δ, λ, B, L$_{SR}$, β$_R$, *Pm*, *Cmmm*, γ (relaxed) and η phases of Ta$_2$O$_5$,with different exchange-correlation potentials and optimized lattice parameters. Some theoretical and experimental values are listed for comparison.

| | This work | | | Literature | | |
|---|---|---|---|---|---|---|
| | LDA-CA | GGA-PBE | GGA-PBEsol | Hybrid functionals | *GW* | Expt. |
| | | | β$_{AL}$ phase | | | |
| DFT | 0.23(*i*) | 0.26(*i*) | 0.25(*i*) | 1.52*[74] | 1.81, | 3.9[71], 4.2[72] |
| DFT-1/2 | 1.81(*i*) | 1.85(*i*) | 1.85(*i*) | ~2[75], 2.45&[76] | 2.42[74] | 4.5[73] |
| | | | | 1.70#[77] | 1.03[78] | |
| | | | δ phase | | | |
| DFT | 1.23(*i*) | 1.12(*i*) | 1.20(*i*) | ~2[75], 2.92&[76] | 2.22[78] | 3.9[71], 4.2[72] |
| DFT-1/2 | 2.79(*i*) | 2.56(*i*) | 2.73(*i*) | 2.87#[77] | | 4.5[73] |
| | | | λ phase | | | |
| DFT | 2.07(*d*) | 2.13(*d*) | 2.10(*d*) | ~4.0[75] | | 3.9[71], 4.2[72] |
| DFT-1/2 | 4.00(*d*) | 4.09(*d*) | 4.06(*d*) | 3.7(*d*)*[12] | | 4.5[73] |
| | | | | 4.03(*d*)[77] | | |
| | | | B phase | | | |
| | | | | | | 3.9[71], 4.2[72] |
| DFT | 3.14(*i*) | 3.13(*i*) | 3.15(*i*) | 4.7(*i*)*[12] | | 4.5[73] |
| DFT-1/2 | 4.87(*i*) | 4.88(*i*) | 4.90(*i*) | | | |
| | | | L$_{SR}$ phase | | | |
| DFT | 1.61(*i*) | 1.86(*i*) | 1.74(*i*) | | 2.96[78] | 3.9[71], 4.2[72] |
| DFT-1/2 | 2.61(*i*) | 2.91(*i*) | 2.82(*i*) | | | 4.5[73] |
| | | | β$_R$ phase | | | |
| DFT | 1.94 (*i*) | 1.98 (*i*) | 1.96 (*i*) | | | 3.9[71], 4.2[72] |
| DFT-1/2 | 3.89 (*i*) | 3.96 (*i*) | 3.94 (*i*) | | | 4.5[73] |
| | | | *Pm* phase | | | |
| DFT | 2.37 (*i*) | 2.38 (*i*) | 2.37 (*i*) | | | 3.9[71], 4.2[72] |
| DFT-1/2 | 4.28 (*i*) | 4.32 (*i*) | 4.31 (*i*) | | | 4.5[73] |
| | | | *Cmmm* phase | | | |
| DFT | 2.09 (*i*) | 2.17 (*i*) | 2.12 (*i*) | | | 3.9[71], 4.2[72] |
| DFT-1/2 | 4.00 (*i*) | 4.10 (*i*) | 4.06 (*i*) | | | 4.5[73] |
| | | | γ phase (relaxed) | | | |
| | | | | | | 3.9[71], 4.2[72] |
| DFT | 2.26(*i*) | 2.32(*i*) | 2.26(*i*) | 3.75, 4.51&[13] | | 4.5[73] |
| DFT-1/2 | 4.17(*i*) | 4.23(*i*) | 4.21(*i*) | | | |
| | | | η phase | | | 3.9[71], 4.2[72] |

| | | | | |
|---|---|---|---|---|
| DFT | 2.22(*i*) | 2.30(*i*) | 2.25(*i*) | 4.5[73] |
| DFT-1/2 | 4.12(*i*) | 4.24(*i*) | 4.20(*i*) | |

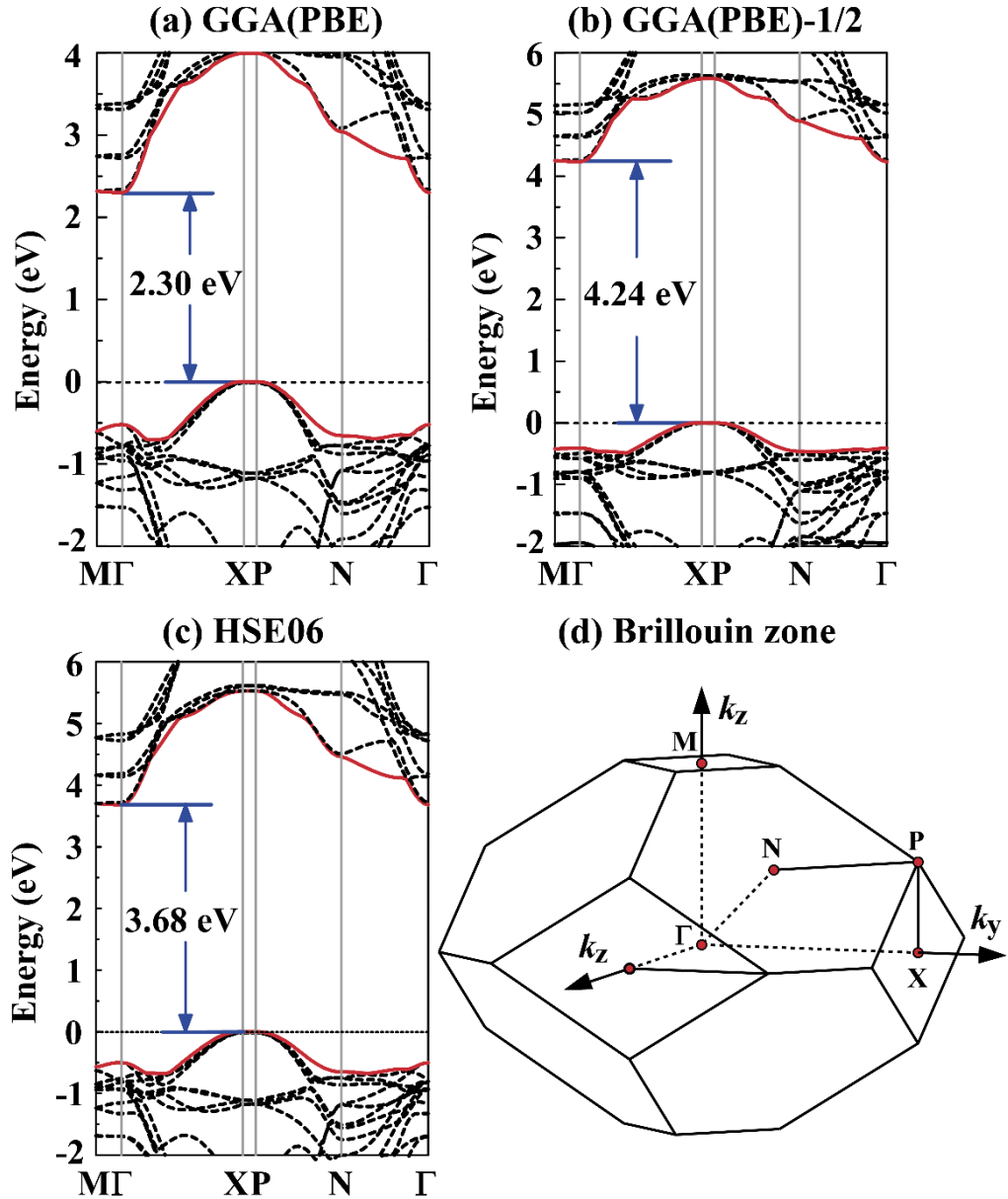

**Figure 7.** (a) Electronic band diagram of η-Ta$_2$O$_5$ calculated using GGA(PBE); (b) electronic band diagram of η-Ta$_2$O$_5$ calculated using GGA(PBE)-1/2; (c) electronic band diagram of η-Ta$_2$O$_5$ calculated using HSE06, with the PBE-optimized structure; (d) the first Brillouin zone of space group *I*4$_1$/*amd*.

The electronic structures of η-Ta$_2$O$_5$ are demonstrated in **Figure 7**, where the band diagrams show indirect (X-Γ) band gaps of 2.30 eV and 4.24 eV according to GGA(PBE) and GGA(PBE)-1/2. The 4.24 eV indirect band gap is a reasonable value that recovers the electronic structure of L-$Ta_2O_5$ fairly well, and superior to the HSE06 band gap value of 3.68 eV, but the computational time of GGA-1/2 is typically less than 1.15 times the standard GGA calculation time for the same supercell.[79] Hence, it is in principle possible to simulate a $Ta_2O_5$-based supercell with more than 500 atoms, without suffering from the band gap underestimation problem. The HSE06 band gap is slightly underestimated because its standard screening length for exact exchange is designed for moderate band gap semiconductors, while it is possible that insufficient exact exchange is mixed for insulators with band gaps greater than 4 eV. However, the morphologies of band structures in **Figure 7(b)** and **7(c)** are quite similar, supporting the validity of GGA-1/2 band structures.

In **Table 2** we list the calculated band gaps for other phases, where the representative GGA(PBE)-1/2 gap values indeed show great variability. For instance, the $\beta_{AL}$ phase only demonstrates an indirect 1.85 eV band gap, followed by the hexagonal δ-phase which also shows a relatively narrow 2.56 eV band gap. On the contrary, the high pressure B-phase owns the largest 4.88 eV band gap. The band gaps calculated for most orthorhombic/monoclinic/tetragonal phases are close to 4 eV, including λ, $\beta_R$, *Pm* and *Cmmm* phases. The low band gap for the $L_{SR}$ phase stems from the existence of oxygen vacancy sites. Finally, compared with reported theoretical values in the literature, we find that the GGA-1/2 method predicts very similar data with the correct sequence of band gaps among various model structures.

## *Band offset between Ta and Ta₂O₅*

As Ta$_2$O$_5$ is an excellent candidate material for memristors, it is worthwhile to investigate its interfacial properties with respect to possible metal electrodes. As an application of the high symmetry η-Ta$_2$O$_5$ structure, we used the GGA-1/2 method to explore the electronic properties of the Ta/Ta$_2$O$_5$ interface. The structures of Ta and Ta$_2$O$_5$ were body-centered cubic and tetragonal *I*4$_1$/*amd*, respectively, but the mismatch of their lattice constants is as large as 17%. Yet, as the *a*/*b* lattice constants of both structures are less than 4 Å, we could reach a relatively good match by combining the $\sqrt{10} \times \sqrt{10}$ *a-b* surface of Ta and the $2\sqrt{2} \times 2\sqrt{2}$ *a-b* surface of Ta$_2$O$_5$, while keeping the total interfacial area for the supercell as low as 1.15 nm$^2$. The fully optimized atomic structures and the layer-decomposed local density of states are demonstrated in **Figure 8**. The contact of Ta/Ta$_2$O$_5$ has been shown to be Ohmic according to our GGA-1/2 calculation, since the Fermi level is aligned with the conduction band edge inside Ta$_2$O$_5$. The exact contact property of Ta/Ta$_2$O$_5$ was rarely reported in the literature, but recently Kim *et al.* assumed that the Ta/Ta$_2$O$_5$ interface points to Ohmic contact,[80] which is consistent with our results.

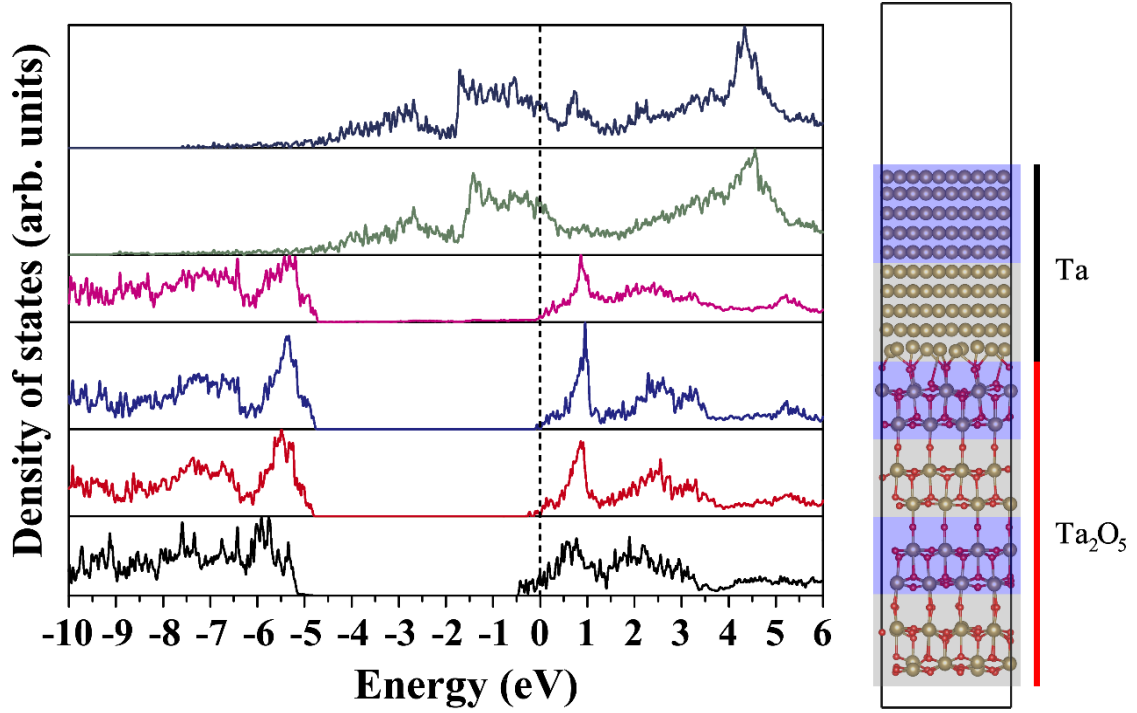

**Figure 8.** Layer-decomposed local density of states (LDOS) of our optimized Ta/Ta2O5 interface model, with the model structure and layer-decomposition scheme illustrated on the right. Each LDOS corresponds to a shaded layer following the same up-down sequence, and the Fermi level is set to zero energy.

## IV. CONCLUSION

In summary, we find that the triclinic γ-phase of $Ta_2O_5$ is transformed into a much more symmetric $I4_1/amd$ structure (η-phase $Ta_2O_5$) during structural relaxation. Compared with various other $Ta_2O_5$ models, η-phase $Ta_2O_5$ has the lowest energy. Hence, both the high temperature α-phase and low temperature η-phase of $Ta_2O_5$ can be assigned the $I4_1/amd$ space group, but the α-phase is dynamically unstable at zero temperature. Using the GGA-1/2 self-energy correction method, the band gap of η-phase $Ta_2O_5$ is shown to be indirect (X-Γ) with a value of 4.24 eV. The high symmetry tetragonal model structure combined with the efficient GGA-1/2 method enables *ab initio* simulation of Ta$_2$O$_5$-based devices. In particular, we confirm the Ohmic contact nature of Ta/Ta$_2$O$_5$ interface according to *ab initio* calculations, taking advantage of the simple structure of η-Ta$_2$O$_5$.

## ACKNOWLEDGEMENTS

This work was financially supported by the MOST of China under Grant No. 2016YFA0203800, the National Natural Science Foundation of China under Grant No. 11704134 and 51732003, and the Fundamental Research Funds of Wuhan City under Grant No. 2017010201010106. L.R.C. Fonseca thanks the Brazilian agency CNPq for financial support under Grant No. 118485/2017-2.